# Quantification of bound water content, interstitial porosity and fracture porosity in the sediments entering the North Sumatra subduction zone from Cation Exchange Capacity and IODP Expedition 362 resistivity data

Jade DUTILLEUL, Sylvain BOURLANGE, Marianne CONIN, Yves GERAUD, Université de Lorraine, CNRS, GeoRessources, F-54000 Nancy, France

**Abstract**

In this study, we investigate porosity evolution through the sedimentary input section of the North Sumatra Subduction zone by quantifying interstitial porosity, bound water content and fracture porosity based on IODP Expedition 362 data and post-cruise chemical analyses. During IODP Expedition 362, total porosity of the sedimentary section entering the North Sumatra subduction zone was measured. This total porosity is derived from the total water content of core samples thus including pore water and water bound to hydrous minerals like smectite. Clay mineral composition varies over the sedimentary section and is mainly kaolinite/illite in the Nicobar Fan units and smectite/illite in the prefan pelagic unit below. The prefan pelagic unit shows anomalously high total porosity values and is stratigraphically correlated to a high amplitude negative polarity (HANP) seismic reflector located landward. This HANP reflector has been previously interpreted as a porous fluid-rich layer where the décollement may develop along parts of the margin as a consequence of pore pressure buildup. We estimate clay bound water content from Cation Exchange Capacity (CEC) which gives information about the smectite/illite composition and soluble chloride content data. Interstitial porosity corresponds to onboard total porosity corrected from clay bound water and is more relevant in terms of sediment compaction state and fluid flow properties. Interstitial porosity versus vertical effective stress curve shows no evidence of undercompaction and suggests that the input section has been experiencing normal consolidation due to high sediment accumulation rate. The porosity anomaly observed in the prefan pelagic unit results from the local occurrence of water-bearing minerals like smectite rather than excess pore pressure, which might, however, buildup more landward in the basin. We also estimate fracture porosity using a resistivity model for shales used in previous works based on wireline resistivity log and show that fracture porosity yields 4-6% in damaged parts of the sedimentary section investigated.

## 1. Introduction

Sediment compaction in accretionary wedge sediments are expected to drive shallow seismogenesis and thus tsunamigenesis (McNeill et al., 2017a). Among a wide range of rock physical and hydrogeological properties, the assessment of the compaction state is principally based on porosity (Bangs et al., 1990; Bray & Karig, 1985; Hart et al., 1995; Screaton et al., 2002; Conin et al., 2011). A typical approach is to identify zones showing abnormal either low or high porosity compared to a reference porosity-effective stress curve assuming hydrostatic fluid pressure and uniaxial compaction conditions (Screaton et al., 2002; Conin et al., 2011 ; Hüpers et al., 2015). Anomalously high porosity zones are typically interpreted as under-compacted zones where excess pore pressure resulting from the ineffectual dewatering of low-permeability marine sediments tends to prevent consolidation whereas anomalously low porosity zones rather indicate compressive tectonic stress and/or erosional unloading (Conin et al., 2011). However, such analysis cannot be directly carried on the porosity routinely measured on core samples during International Ocean Drilling Program (IODP) Expeditions, which is based on measurement of the total water content including both the volume of free water contained in pores corresponding to interstitial porosity and the volume of water bound to hydrous minerals such as clay, opal or zeolites (Brown & Ransom, 1996; Henry, 1997 ; Henry & Bourlange, 2004) referred here to as bound water content. Bound water corresponds to interlayer water and water adsorbed on mineral external surfaces (Henry & Bourlange, 2004).

Bound water can be expelled from the mineral structure by mineral dehydration reactions that essentially depends on time and temperature (Moore & Vrolijk, 1992). Contrary to interstitial water, only a limited amount of smectite-bound water is thought to be expelled from the mineral structure by compaction (Bird 1984; Colten-Bradley, 1987; Fitts & Brown, 1999; Henry & Bourlange, 2004). Hence, compaction state analysis should be based on interstitial porosity. Such correction to estimate interstitial porosity have been applied previously on core samples from Barbados and Nankai accretionary margins using different methods. The first method is based upon smectite content determined from XRD (Brown & Ransom, 1996; Hashimoto et al., 2010) while the second depends on Cation Exchange Capacity (CEC) measurement (Henry, 1997; Henry & Bourlange, 2004; Conin et al., 2011).

In 2016, IODP Expedition 362 drilled two sites U1480 and U1481 in the basin entering the North Sumatra Subduction zone, ~225 kilometers seaward of the deformation front, in the southern portion of the 2004 9.2 Mw earthquake rupture zone (McNeill et al., 2017a). The input sedimentary section, composed of the distal part of the trench wedge, the Bengal-Nicobar fan siliciclastic sequence and a prefan pelagic clay-rich interval (McNeill et al., 2017b), as well as the topmost basalt of the 60-70 Ma old oceanic crust were logged and sampled. Porosity measured onboard on core samples is total porosity and does not provide any insight in the contribution of moveable fluid, irreducible bound fluids (clay bound water and capillary bound water) and fractures to *in situ* properties of the input sedimentary section without further analysis. We use IODP Expedition 362 total porosity and wireline resistivity data, as well as post-cruise chemical analyses including Cation Exchange Capacity (CEC), soluble chloride content and exchangeable cation composition to quantify bound water content, interstitial porosity and fracture porosity. CEC is measured on natural samples and is an average value over all the minerals of the samples. It is used as a proxy for clay mineral composition in the entering sedimentary section at Sites U1480 and U1481. CEC and soluble chloride content are used to estimate the volume of clay bound water. Onboard total porosity data is corrected from clay bound water to quantify interstitial porosity at the two sites. Resulting porosity versus effective stress curves are interpreted in terms of compaction state and origin of the apparent porosity anomalies. Because total porosity is measured onboard on undeformed samples, it does not take into account fracture porosity, contrary to wireline resistivity. We estimate fracture porosity at Site U1481 using a resistivity model for shales (Revil et al., 1998) accounting for pore fluid composition, exchangeable cation composition and temperature and interpret the results regarding faults, fractures and drilling induced deformation observed on cores.

## 2. Geological background and drilled sites location

During IODP Expedition 362, drilled Sites U1480 and U1481 (Figure 1) revealed a complete sedimentary succession from Late Cretaceous to Pleistocene composed of the distal part of the trench wedge, the Bengal-Nicobar fan succession, and the prefan pelagic succession on basaltic crust (McNeill et al., 2017a and b). The sediments are generally unlithified to partially lithified but some lithified intervals were found near basement. The input section is particularly thick, being ~1.5 km at the drill sites and reaching 4-5 kilometers at the deformation front. It is mainly composed by the Himalayan-derived Bengal-Nicobar fan that developed during the Miocene with a significant increase in sediment flux in the 9.5-2 Ma interval (McNeill et al., 2017b). In this respect, the North Sumatra margin differs from other well studied subduction margins and is used as a proxy for thickly sedimented but poorly sampled subduction margins with unknown hazards potential such as the southern Lesser Antilles and the Makran (McNeill et al., 2017a).

Site U1480 and Site U1481 provided samples from seafloor to ~1432 mbsf (meters below seafloor) and from ~1150 mbsf to ~1499 mbsf respectively. Lithologic Units I to VI were defined at Site U1480 (McNeill et al., 2017a) (Figure 2). Units I and II represent Nicobar Fan sedimentation, Units III-V represent pelagic pre-fan sedimentation and magmatic intrusion and Unit VI is basaltic oceanic crust with moderate to high alteration. Unit I (0 - ~26 mbsf) is calcareous clay and silty clay with occasional ash and alternating fine-grained sand and clay. Unit II (~26 - ~1250 mbsf) is characterized by detrital silt and sand and subdivided into Subunits IIA (~26 - ~344 mbsf), IIB (~344 - ~786 mbsf) and IIC (~786 - ~1250 mbsf). Subunit IIA is sandy silt and fine-grained sand, with silty clay and silt. Subunit IIB is alternating bed of laminated silt and clay. Subunit IIC is gray or black clay/claystone,silty clay/claystone and structureless muddy sand/sandstone. Clay-rich Unit III (~1250 - ~1327 mbsf) is more lithified than Unit II and subdivided into Subunits IIIA (~1250 - ~1310 mbsf) and IIIB (~1310 - ~1327 mbsf). Subunit IIIA is poorly diagenetised gray-green claystone with foraminifers whereas Subunit IIIB is brown-red tuffaceous and silty claystone with biosiliceous debris such as fragmented sponge spicules, foraminifers and radiolaria (Hüpers et al., 2017), palagonite and minor chalk. Unit III shows apparent porosity and freshening anomalies (Hüpers et al., 2017) (Figure 2) and is stratigraphically equivalent to a high amplitude negative polarity (HANP) seismic reflector interpreted landward as a porous fluid-rich layer weaker than the overlying sediments and inferred to be the locus for initiating the décollement fault (Dean et al., 2010). Unit IV is basaltic flow overlying tuffaceous sandstones/conglomerates and volcanic breccia. Unit V is calcareous claystone and chalk with intercalated magmatic intrusions. Site U1481 has sampled the lower part of Unit II (~1150 - ~1360 mbsf) and the upper part of Unit III (~1360 mbsf - ~1499 mbsf) (Figure 3). No subunits have been identified but Unit II correlates to Subunit IIC at Site U1480 and Unit III correlates to Subunit IIIA based on biostratigraphic and lithologic analyses (McNeill et al., 2017a).

## 3. Data and methods

### 3.1 Onboard acquired data

XRD data acquired onboard shows clay content with significant variation through the input sedimentary section. Mean clay content increases with depth through the Nicobar fan units and the siliciclastic prefan units (Figure 2). At Site U1480, mean clay content is ~55% in Unit I,

~48% in Subunit IIA, ~58% in Subunit IIB, ~60% in Subunit IIC, ~70% in Subunit IIIA, ~76% in Subunit IIIB (chalk excluded) and ~68% in the tuffaceous part of Unit IV. This reflects the different stages of fan development, with clay-rich distal deposits during fan initiation followed by lower and middle fan successions with higher quartz and lower clay contents (McNeill et al., 2017a). Prefan Units III-V represents more slowly accumulated, nearly sandstone-free, pelagic and hemipelagic prefan succession with less quartz and higher clay content, that shows evidences of locally-derived material resulting from the erosion of previously deposited sedimentary rocks or magmatic rocks and syndepositional reworking (McNeill et al., 2017a). Mean clay content is higher in Subunit IIIA at Site U1480 than in the equivalent Unit III at Site U1481 where it is ~63% (Figure 3). Clay content also shows a wide range of variations depending on the lithology. At Site U1480, in Nicobar fan Units I and II, clay content scatters from ~11% to ~69% in sandstones (~35% in average), from ~30% to ~71% in calcareous claystones (~47% in average), and from ~36% to ~73% in siliciclastic claystones (63% in average). At Site U1481, in Unit II, it ranges from ~27% to ~49% in sandstones and from ~54% to ~72% in siliciclastic claystones. XRD patterns indicate the occurrence of kaolinite/chlorite and illite in Unit II and smectite in Unit III (McNeill et al., 2017a).

Total porosity is derived from Moisture and Density (MAD) measurements carried onboard following standard procedure (Blum, 1997). First, the wet weight of the saturated sample is measured and its wet volume is measured using a helium pycnometer. The sample is then dried in a convection oven at 105°C ± 5°C for 24 hours (Blum, 1997). The sample is finally cooled to room temperature in a desiccator so that dry weight and dry volume can be measured 2 hours later. The calculation of total porosity from wet and dry masses and either wet or dry volume accounts for the precipitation of salt during drying. Globally, total porosity of mud- and sand-rich samples decreases with depth at Site U1480 (Figure 2) and at Site U1481 (Figure 3), except in Unit III where mud samples locally exhibit a positive shift of total porosity values. Evolution of total porosity is developed in the Results section.

Electrical resistivity was measured each 0.15 m at Site U1481 from ~730 mbsf to 1484 mbsf using the wireline High-Resolution Laterolog Array (HRLA) Schlumberger tool (McNeill et al., 2017a). At Site U1480, due to operating issues, electrical resistivity was recorded on less than 60 meters. The HRLA provides an array of five apparent resistivities (RLA1-5), each with increasing depth of investigation. RLA1-5 were used by a radius inversion algorithm to compute the true resistivity of the formation (McNeill et al., 2017a). True resistivity exhibits a relatively constant trend in Unit II where it scatters from 1.2 to 4.5 ohm.m (2.0 in average) mbsf (McNeill et al., 2017a). It is less scattered in Unit III where it shows a minimum ~1.4 ohm.m near 1400 mbsf (McNeill et al., 2017a).

**3.2 On shore acquired data**

Chemical analyses were carried on at the Laboratoire Interdisciplinaire des Environnements Continentaux (LIEC) in Nancy and Metz, France. CEC was determined by exchange with cobaltihexamine and ultraviolet-visible spectrometer Varian SpectrAA 800 Zeeman. Soluble chloride content per dry mass was determined by sequential water extraction (Tessier et al., 1979) and ion chromatography. Exchangeable cation composition ($Na^+$, $K^+$, $Ca^{2+}$ and $Mg^{2+}$) was measured by atomic absorption spectrometer Thermo Scientific ICE 3300. When possible, CEC, soluble chloride content and exchangeable cation composition were measured both on a core sample and a near-by squeeze cake, which is a sample residue of onboard pore fluid extraction under pressure. At Site U1480, CEC and soluble chloride content analyses were carried on 57 core samples and 61 squeeze cakes. At Site U1481, CEC, soluble chloride content and exchangeable cation composition were measured on 23 core samples and 22 squeeze cakes. The sampling interval is ~25 m and ~16 m in average at Site U1480 and Site U1481 respectively. Pore water chemical composition was analyzed onboard from fluid extracted under pressure.

- **Estimation of bound and interstitial porosities from CEC data**

Two types of water are typically distinguished in clay-rich sediments saturated by a saline fluid. Interstitial water refers to the pore water whose chemical composition is close to that of sea water (i.e.: chloride bearing). In contrast, clay-bound water is chloride-free and has an organized structure. In smectite-rich sediments, bound water occurs in interlayer space and adsorbed on external surfaces as a result of the compensation of negatively charged smectite layers by hydrated cations such as $Na^+$, $K^+$, $Ca^{2+}$ and $Mg^{2+}$. Smectite commonly shows three hydration states with specific amount of interlayer water and thus interlayer thickness (Colten-Bradley, 1987) that depends upon the smectite layer charge and the type of cation (Ransom & Helgeson, 1995). Interlayer water is organized in one, two or three water layers, which typically corresponds to 6-8, 12-16 and 18-24 water molecules per cation charge (Henry, 1997) and can reach 25% of the total volume of water (Henry & Bourlange, 2004; Conin et al., 2011). Interstitial porosity can thus be estimated by correcting total porosity from clay bound water either using CEC (Henry, 1997) which indicates the maximum number of surface exchangeable cations per unit mass of sediment (Revil et al., 1998) or the mass of smectite interlayer water estimated from smectite hydration state and smectite wt% in the bulk sediment (Brown & Ransom, 1996).

According to the correction of total porosity based on CEC data (in mol/kg of dried sample), interstitial porosity ($\phi_i$) is expressed :

$$\phi_i = \phi_t - \phi_b = \phi_t - n \frac{M_w}{\rho_w} CEC \rho_g (1 - \phi_t) \qquad (1)$$

where $\phi_t$ is MAD total porosity measured onboard on core samples, $\phi_b$ is bound water content, $M_w$ and $\rho_w$ are respectively water molar mass and density ($M_w$ = 0.018kg/mol; $\rho_w$ = 1024 kg/m³), $\rho_g$ is the grain density (in kg/m³) and $n$ is the average number of water molecules per cation charge estimated by the relationship between bound water ratio $w_b$ and CEC as determined empirically on samples from different subduction zone locations (Henry & Bourlange, 2004). Since bound water is assumed chloride free contrary to interstitial water, $w_b$ is expressed :

$$w_b = \frac{m_w}{m_g} - \frac{m_{w,Cl}}{m_g} = \frac{m_w}{m_g} - \frac{C_s \rho_w}{[Cl^-]} \qquad (2)$$

where $m_g$ is the mass of grains, $m_w$ is the total mass of water and $m_{w,Cl}$ is the mass of chloride bearing water, determined experimentally from $C_s$ the soluble chloride concentration per dry mass (in mol/kg), [Cl⁻] the chloride concentration in the pore fluid extracted under pressure (in mol/l). Previous works using samples from Nankai Site 1173 (Henry & Bourlange, 2004) and Site C0001 (Conin et al., 2011), and Barbados (Henry, 1997) find $n$ ranging from 12 to 15 (Figure 4) despite data with CEC < 0.20 mol/kg, which exhibit a wider range of variation.

- **Estimation of fracture porosity from CEC, exchangeable cation composition and resistivity data**

Formation bulk conductivity $\sigma_b$ can be modeled as the sum of $\sigma_{if}$ the conductivity of the interstitial fluid-filled fractures and $\sigma_r$ the conductivity of rock fragments weighted by a relative volume coefficient $\phi_c$ for the fracture and $1 - \phi$ for the rock fragments (Bourlange & Henry, 2003) :

$$\sigma_b = \phi_c \sigma_{if} + (1 - \phi_c) \sigma_r \qquad (3)$$

Hence, minimum fracture porosity can be estimated knowing $\sigma_b$, $\sigma_{if}$ and $\sigma_r$ :

$$\phi = \frac{\sigma_b - \sigma_r}{\sigma_{if} - \sigma_r} \qquad (4)$$

$\sigma_r$ is estimated using a resistivity model for clay (Bourlange et al., 2003; Conin et al., 2011) based on Revil et al.'s model (1998). This model is based on Archie's law (Archie, 1942) linking the resistivity-derived porosity $\phi$ to the formation factor $F$ :

$$F = a\,\phi^{-m} \quad (5)$$

where $m$ is the cementation factor typically ranging from 1 to 3.5 (Conin et al., 2011) and $a$ is a constant. Archie (1942) originally proposed that $a = 1$ allows a satisfying fit between equation (1) and sand data, but core measurements typically shows that good fitting requires $a \neq 1$ (Carothers, 1968; Porter & Carothers, 1970; Gomez-Rivero, 1977).

Formation factor $F$ is defined as the ratio of interstitial fluid conductivity $\sigma_{if}$ to bulk conductivity $\sigma_b$. Clay-rich sediments such as hemipelagites from Nankai (Bourlange & Henry, 2003) or Sumatra (Figures 2 and 3) have elevated total clay content and thus a very high surface conductivity $\sigma_s$. In that case, $F$ is defined as the limit of $\frac{\sigma_{if}}{\sigma_b}$ for a zero surface conductivity (Bussian, 1983) :

$$\frac{1}{F} = \lim_{\sigma_s \to 0} \frac{\sigma_b}{\sigma_{if}} = \frac{\phi^m}{a} \quad (6)$$

with $a = 1$ for theoretical consistency (Glover et al., 2000; Bussian, 1983). Several studies (Roberts & Schwartz, 1985; Schwartz & Kimminau, 1987; Sen, 1987) have confirmed this law (Revil et al., 1998).

As suggested by Bourlange et al. (2003), the Formation Factor can also be calculated using a high-salinity asymptotic approximation of equation (10) in Revil et al. (1998) obtained through a Taylor expansion calculation, valid for $\xi = \frac{\sigma_s}{\sigma_{if}} \ll 1$

$$F = \frac{\sigma_{if}}{\sigma}\left[1 + 2\xi\left(\frac{\sigma_{if}}{\sigma} - 1\right)\right] \quad (7)$$

Contrary to MAD total porosity $\phi_t$ measured at discrete depths, resistivity-derived porosity $\phi$ can be calculated at any depth where wireline resistivity was acquired. Exchangeable cation composition, CEC and interstitial fluid conductivity $\sigma_{if}$ are linearly extrapolated between the 45 samples analyzed at Site U1481. Equation (6) with $a = 1$ and $m$ between 1 and 3.5 is combined with equation (7) considering that $\sigma$ is $\sigma_b$ the bulk conductivity. Reciprocally, $\sigma_r$ is obtained by solving equation (7) assuming that $\sigma$ is $\sigma_r$ and thus using $\phi_t$ instead of $\phi$ in equation (6). $\sigma_b$, $\sigma_{if}$ and $\sigma_s$ are calculated with the following analytical expressions (Revil et al., 1998; Bourlange, 2003).

Bulk conductivity $\sigma_b$ is the inverse of the true formation resistivity measured with the HRLA tool.

Interstitial fluid conductivity $\sigma_{if}$ is calculated from pore fluid composition determined onboard, applying a linear correction at any depth since it differs from seawater composition :

$$\sigma_{if} = \sigma_{sw} \frac{\sum_i \left(\beta_f^i \times Z_i \times C_{iws}^i\right)}{\sum_j \left(\beta_f^j \times Z_j \times C_{sw}^j\right)} \quad (8)$$

where $C_{iws}^i$ (respectively $C_{sw}^i$) is the concentration determined on squeezed samples (respectively sea water) for $Cl^-$, $Na^+$, $K^+$, $Ca^{2+}$, $Mg^{2+}$, $SO_4^{2-}$, $\beta_f^i$ is the ionic mobility in the fluid (Revil et al., 1998), $Z_i$ is the number of charges of ions and $\sigma_{sw}$ is the sea water conductivity which depends on the temperature $T$ as :

$$\sigma_{sw} = 5.32(1 + 0.02(T - 25)) \quad (9)$$

With temperature $T$ function of depth $z$ (m) as :

$$T(°C) = 1.79 + 44.4 \times 10^{-3} z \quad (10)$$

obtained from thermal conductivity measurement on cores and assuming a heat flow of 72.6 mW/m² (Hüpers et al., 2017). This heat flow was determined by least square fitting from the extrapolation of formation temperature measured in the first ~200 meters at Site U1480.

Considering a medium constituted of insulating grains coated by a conductive layer, surface conductivity $\sigma_s$ is calculated from CEC, grain density $\rho_g$ and equivalent surface mobility $\beta_s$ (Bourlange et al., 2003 modified from equations (7) and (8) in Revil et al., 1998) :

$$\sigma_s = \frac{2}{3} \rho_g \, CEC \, \beta_s \qquad (11)$$

Equivalent surface mobility $\beta_s$ is a linear combination of $Na^+$, $K^+$, $Ca^{2+}$, $Mg^{2+}$ exchangeable cation surface mobility $\beta_s^i$, charge $Z_i$ and concentration in the surface layer $C_i$ :

$$\beta_s = \frac{\sum_i(\beta_s^i \times Z_i \times C_i)}{\sum_j(Z_j \times C_j)} \qquad (12)$$

A linear temperature dependency of the exchangeable cation surface mobility $\beta_s^i$ is assumed :

$$\beta_s^i = \beta_s^{i*}(1 + \vartheta_s^{i*}(T\text{-}25)) \qquad (13)$$

Parameters $\beta_s^{i*}$ and $\vartheta_s^{i*}$ are reviewed by Revil et al. (1998).

## 4. Results

### 4.1. CEC, Bound water ratio and lithology

At Sumatra sites, CEC measured on core samples ranges between 0.04 and 0.63 mol/kg, reaching values higher than at any site in any Nankai transects (Henry & Bourlange, 2004; Underwood & Steurer, 2003; Conin et al., 2011) but lower than Barbados accretionary wedge highest values (Henry, 1997) (Figure 4). CEC content shows a strong dependence on lithology and significantly differs between the Nicobar fan sequence Units I and II and prefan Units III, IV and V. Data from Units II and III at Site U1481 are similar with data from equivalent Subunit IIC and Subunit IIIA at Site U1480. In Units I and II, CEC values correlate well with total clay content (Figure 5), remaining below 0.25 mol/kg at Site U1480 and 0.35 mol/kg at site U1481. At both sites, CEC content shows a wide dispersion through Unit III between ~0,18 and ~0,63 mol/kg where total clay content ranges between 40% to 85%. The highest CEC values (> 0.5 mol/kg) are observed in the lowermost part of clay-rich Subunit IIIA, in the clay-rich sample with ashes of Subunit IIIB and in the tuffaceous sample of Unit IV at Site U1480. In Unit V, some calcareous claystones show high CEC values whereas CEC is ~0 for nearly pure carbonate samples. This range of CEC values suggests that clay minerals are mainly illite and/or kaolinite and/or chlorite in Units I and II, and smectite/illite in Unit III. This result is in accordance with onboard XRD analysis (McNeill et al., 2017a) and with the temperature at Site U1480 which is estimated around 45-80°C in Subunit IIIA assuming an uncertainty of ~20% on thermal conductivity measurements (see also Hüpers et al., 2017) since only limited illitization is expected at such temperature.

Bound water ratio of clay-rich samples from Sumatra Sites U1480 and U1481 is plotted versus CEC in Figure 4 and compared to that of clay-rich sediments from Barbados and Nankai determined in previous works. Bound water ratio for Sumatra samples with CEC values less than 0.2 mol/kg is significantly scattered, as observed in lower proportion for sediments from Nankai Upper Shikoku Basin and Kumano Basin Site C0001. On the contrary, bound water ratio for Sumatran samples with CEC values above 0.2 mol/kg, as well as samples from Nankai Lower Shikoku Basin and Barbados are less dispersed and plot on a $n = 15$ trend for Sumatra samples, higher than the $n = 12$ trend determined for Nankai Lower Shikoku Basin (Henry & Bourlange, 2004) and Barbados samples (Henry, 1997). These values correspond to smectite minerals with two water layers. 15 water molecules per cation charge is the number found by Ransom & Helgeson (1994) for an idealized smectite and by Conin et al. (2011) for samples at Nankai Kumano Basin Site C0001. As suggested by Henry and Bourlange (2004), this difference of behavior may be related to the lithology of the samples analyzed. Samples displaying a wide range of bound water ratio reflect presence of heterogeneous lithologies : alternating volcanic ash and hemipelagic clays with more or less biogenic silica for Nankai Upper Shikoku Basin facies (Henry & Bourlange, 2004), silty clay and hemipelagites from slope apron and accreted fine-grained terrigenous material and hemipelagites for samples from Nankai Kumano Basin Site C0001 (Conin et al., 2011) and the fine-grained silicilastic Nicobar fan sequence for Sumatran Unit I and Unit II samples (Figure 6) (McNeill et al., 2017a). For such samples with both low CEC and chloride free water content, it is generally difficult to estimate $n$ by plotting bound water ratio versus CEC (Henry & Bourlange, 2004). In contrast, samples plotting on the same trend such as samples from Unit III at Sites U1480 and U1481 (Figure 6) and samples from Nankai Lower Shikoku Basin and Barbados are relatively homogeneous claystones (McNeill et al., 2017a; Underwood & Deng, 1997). Bound water ratio also seems to increase with total clay content and presumably smectite content. Through Subunits IIB and IIC, Units III and IV, Sumatra samples show total clay content, smectite content and bound water ratio that become progressively greater whereas Unit I and Subunit IIA exhibit lower total clay and smectite content (Figure 6).

### 4.2. Porosity corrections

- **Total porosity**

At Site U1480 (Figure 2), total porosity of mud samples is very high in the very shallow Unit I (~43% up to ~80%, average ~66%), and exponentially decreases through Unit II following the total porosity-vertical effective stress curve $\phi_t = 50.6 e^{-0.044\,\sigma'_v}$ (obtained by non-linear least squares fitting with R² = 0.7410) from ~34% - 76% in Subunit IIA (average, 48%), to ~21% - 48% in Subunit IIB (average, 39%), to ~11% - 44% (average, 33%) in Subunit IIC. In the lower part of Subunit IIIA, between ~1265 and ~1310 mbsf ($\sigma'_v$~ 13.2-13.7 MPa), porosity values of mud samples display a wide scatter, ranging from 26% up to 44% (in average, 31%) resulting in a sharp positive porosity anomaly of 16% compared to the reference total porosity-vertical effective stress curve. At Site U1481, where only a limited portion of the sedimentary sequence was sampled and logged, we used the total porosity-vertical effective stress curve determined at Site U1480 (Figure 7). This curve fits very well to the total porosity of mud samples from Site U1481 which decreases from 27% - 35% (average, 31%) in the investigated portion of Unit II to 23% - 32 % (average, 28%) in topmost Unit III. Between ~1375 and ~1430 mbsf ($\sigma'_v$ ~14.5 – 15.2 MPa), total porosity shifts from ~28% to a maximum of ~32%, resulting in a wide positive anomaly with relatively low amplitude and poor dispersion.

- **Interstitial porosity and bound water content**

At Site U1480, interstitial porosity shows an exponential decrease with vertical effective stress $\phi_i = 42.2 e^{-0.053\, \sigma'_v}$ (R² = 0.6288) through Unit I to Unit IV (Figure 7). This curve is also valid at Site U1481 where less data are available. At Site U1480, bound water content varies from ~4% up to ~13% (in average, 7%) in Nicobar fan sequence Units I and II, increases substantially through clay-rich Unit III up to 25%, and decreases to ~3% in calcareous claystones of Unit V. At Site U1481, bound water content decreases through the lower part of Unit II investigated from ~17% in the topmost sample analyzed to ~5% near the boundary between Units II and III. In the upper part of Unit III investigated, bound water content averages ~13% which is twice higher that at Unit II and scatters from ~8% up to 25%. Within Unit III, bound water content shows two local increases at $\sigma'_v$ ~ 14.5 and $\sigma'_v$ ~15.2 MPa which corresponds respectively to the top and the base of the high total porosity zone where it yields the highest values ~20%. Only a few data is available in-between and exhibit average bound water content. At both sites, bound water content variations are well correlated to that of CEC content and show substantial increase in the zones showing anomalously high total porosity.

It is known that part of the bound water can be expelled under increasing effective normal stress (Bird 1984; Colten-Bradley,1987; Fitts & Brown, 1999; Henry & Bourlange, 2004). Here, samples have been submitted to vertical effective stress up to ~16 MPa due to burial. In addition, pore-water extraction residues have been squeezed at maximum pressures ~24.5 MPa (McNeill et al., 2017a). It has been shown that the squeezing process drives off part of the bound water (Fitts & Brown, 1999 ; Henry & Bourlange, 2004). This is also suggested by the lower CEC measured on squeeze cakes than on near-by core samples at Sites U1480 and U1481 (Figure 8). The difference in CEC observed between squeeze cakes and core samples depends on lithology and presumably on clay mineral composition. In Units I and II that show low to intermediate CEC and that presumably mainly contain illite and/or kaolinite, CEC is in average 0.03 mol/kg lower in squeeze cakes than in core samples at Site U1480. In smectite-rich Unit III where CEC is higher, CEC can be 0.2 mol/kg higher in the core samples than in the near-by squeeze cake, and reciprocally. Assuming that the squeezing process releases 3-4 water molecules (Henry & Bourlange, 2004) (i.e : $n = 12$ in equation (1)) and fully accounts for the variation in CEC observed between core and squeeze samples, the expelled volume of bound water corresponds to less than 1% (resp. 7%) of the total volume of the samples in Unit I and II (resp. Unit III). However, we assume that this volume is much lower in Unit III because the scatter in CEC values in this unit may be triggered by variations in smectite content, thus it is unlikely that the difference in CEC content between core samples and nearby squeeze cakes fully results from the squeezing process.

- **Resistivity-porosity**

A satisfying fit of resistivity-porosity to Site U1481 MAD total porosity data ($\phi_t$) can be achieved with $a = 1$ and $m = 2.5$. These values are in accordance with previous studies at Nankai accretionary wedge (Bourlange & Henry, 2003; Conin et al., 2011). However, resistivity-porosity substantially exceeds total porosity in three zones (Figure 9a) where fracture porosity derived from total porosity measured on cores and resistivity log data yields 4-6% (Figure 9b). The first zone is located in Unit II between at least 1150 m where the first total porosity measurements are available and ~1170 mbsf whereas the two others occur in Unit III between ~1370 mbsf and ~1415 mbsf and between ~1425 mbsf and ~1450 mbsf.

## 5. Discussion

### 5.1 Origin of the apparent porosity anomaly observed in clay-rich sediments of Unit III

Zones of anomalously high total porosity in clay-rich sediments of Subunit IIIA at Sites U1480 and Unit III at Site U1481 (which is equivalent to Subunit IIIA) are correlated to local increase in CEC content and in pore water silica that are used as proxies for smectite content and biogenic opal respectively. As CEC, pore water silica increase is significantly lower at Site U1481 than at Site U1480. Pore water silica rises from 40 mM to 648 mM at Site U1480 (Figure 2) and from 44 mM to 136 mM at Site U1481 (Figure 3). Unit III is composed of amorphous silica (at least 20 wt% at Site U1480 according to Hüpers et al., 2017) and smectite. The substantial proportion of hydrated minerals in Unit III results in a local elevation of bound water content in the high total porosity zone which is clearly observed at Site U1480 (Figure 7). Since the porosity anomaly is not maintained at both sites after applying the correction accounting for smectite bound water (i.e. : interstitial porosity), the occurrence of smectite with two water layers in high proportion in Unit III is expected to be essentially responsible for the anomalously high total porosity observed in these zones. However, this correction does not account for the effect of hydrated biogenic opal, thus bound water content might be underestimated and interstitial porosity overestimated.

According to interstitial porosity – effective vertical stress curve that globally exhibits mono-exponential decrease, the input sedimentary section has only recorded normal consolidation at this stage. This is in agreement with previous studies assuming uniaxial vertical stress at a reference undeformed site seaward of the deformation trench (e.g. Screaton et al., 2002).

**5.2 Implications for fluid production potential**

Present data suggest that the total porosity anomalies evidenced in clay-rich zones in Unit III results from a local change in clay mineralogy rather than the buildup of excess pore fluid pressure. However, these zones show similar characteristics to that of overpressured clay-rich layers interpreted as décollement or proto-décollement at other well studied subduction zones. At part of Nankai, Barbados, Cascadia and Costa Rica, previous works evidenced clay-rich zones showing anomalously high porosity that are well correlated with a negative polarity seismic reflection (e.g. Shipley et al., 1994, Moore et al., 1998; Bangs et al., 1999 for Barbados, Bangs et al., 2004, Moore & Shipley, 1993 and Mikada et al., 2005 for Nankai Muroto transect and Cochrane et al., 1994 for Cascadia) interpreted as indicating an elevated fluid content (Dean et al., 2010) and/or with freshening anomalies (e.g. Bekins et al., 1995 and Henry, 2000 for Barbados, Henry & Bourlange, 2004 for Nankai, Spinelli et al., 2006 for Costa Rica) that suggest the release of bound water during smectite to illite and/or opal to quartz transitions. Among other processes such as mechanical compaction that occurs relatively seaward, such mineral dehydrations may be an important source of fluids that may generate excess pore fluid pressure in poorly drained conditions (Moore & Vrolijk, 1992). At Sumatra North subduction zone, the clay-rich zones showing anomalously high total porosity in Unit III are correlated with the stratigraphic extrapolation of the HANP reflection which exists nearer to the accretionary wedge, interpreted there as a highly porous and permeable fluid-rich layer, presumably the proto-décollement (Dean et al., 2010; Geersen et al., 2013). They are also correlated with a sharp decrease in pore water chloride content from 572 mM to 521 mM at Site U1480 (Figure 2) and from 566 mM to 516 mM at Site U1481 (Figure 3). This freshening anomaly and the temperature estimated between 45°C and 80°C in Unit III suggest that smectite to illite and/or opal to quartz dehydrations may have started, at least more landward. If this is correct, Unit III exhibits significant potential for fluid production. Existing model (Hüpers et al., 2017) predicts that opal dehydration explains the freshening anomaly observed at Site U1480 and that ~2/3 of biogenic opal has been transformed to quartz. If smectite dehydration reaction is also occurring, reaction progress is relatively low since CEC content indicates that smectite remains relatively abundant at the two sites.

**5.3 Interpretation of fracture porosity**

At Sites U1481 and U1480, porosity data (Figures 9a,b and 10a) can be compared with onboard structural analysis (Figures 9c and 10b), although there was uncertainty in defining structures observed on cores as either natural or drilling induced (McNeill et al., 2017a). At both Sites U1480 and U1481, drilling induced deformation is pervasive but seems to develop preferentially as biscuits and fractures in Unit III and/or zones showing fracture porosity.

The three zones exhibiting fracture porosity at Site U1481 (Figure 9b) are correlated to part of cores were substantial natural and/or drilling induced deformation occurred (Figure 9c) : 1150-1170 mbsf in Unit II and 1370-1415 mbsf and 1425-1450 mbsf in Unit III. Minimum fracture porosity in these zones is ~4-6% which is in accordance with similar estimation in Nankai (Bourlange & Henry, 2003; Conin et al., 2011). The 1370-1415 mbsf interval where a significant rise in the frequency of drilling-induced fractures occurs and resistivity is very low corresponds to the anomalously high total porosity zone in Unit III (Figure 9a).

At Site U1480, fracture porosity was not determined since no resistivity log data are available. As observed at Site U1481, drilling-induced deformation is pervasive in the sedimentary sequence but is more developed in Unit III that shows high total porosity. Two structural domains have been evidenced onboard based on the occurrence of natural deformation (McNeill et al., 2017a). In both domains, fault displacement ranges from a few millimeters to 1-2 cm and sand and mud injections are present indicating either fluid pressure or density contrast between sand and mud. The first domain is poorly faulted and fractured and corresponds to Unit II and the top of Subunit IIIA. In contrast, the second domain, corresponding to the bottom of Subunit IIIA, Subunit IIIB Units IV and V, comprises three times more natural faults and numerous fractures (Figure 10b). In particular, Subunit IIIB shows the highest fault (resp. fracture) frequency with clusters reaching ~ 65 faults (resp. ~55 fractures) per 10 meters. These fractures are stylolitic fractures in Subunit IIIB and joint fractures in Unit V. The occurrence of sealed fractures in Unit III may trigger poorly drained conditions that might be favorable to the increase of pore fluid pressure in case of sufficient in situ fluid production.

## 6. Conclusion

Following previous studies, we show that correcting total porosity measured on cores from water bound to hydrous minerals is critical to assess sediment compaction state. CEC content in the sedimentary section entering North Sumatra subduction zone at Site U1480 and U1481 is moderate in the Nicobar fan sequence (~0.15 mol/kg in average) and relatively elevated in prefan pelagic Unit III (0.18-0.63 mol/kg), indicating a change in clay mineral composition from kaolinite/illite in the Nicobar fan sequence to illite/smectite in Unit III. This results in a sharp increase in bound water content up to ~25% in Unit III, implying that the anomalously high porosity measured on Unit III cores is mainly an artifact from smectite bound water. Compaction profile from interstitial porosity suggests that at ~225 km seaward of the deformation front, the thick input sedimentary section has been experiencing normal consolidation. In particular, pelagic Unit III shows no evidence of undercompaction although constituting a reservoir of mineral bound water and being stratigraphically correlated to a HANP seismic reflector interpreted landward as an overpressured fluid-rich layer potentially acting as the proto-décollement/décollement. The comparison of wireline resistivity data and core data evidences zones where fracture porosity yields 4-6% corresponding to natural faults and fractures or drilling induced deformation.

## 7. Acknowledgements and Data

This research used data provided by the International Ocean Discovery Program (IODP) and freely available on the LIMS Report Interface Page at *web.iodp.tamu.edu/LORE* or on the log database at *mlp.ldeo.columbia.edu/logdb/scientific_ocean_drilling.* CEC, exchangeable cation composition, soluble chloride content, interstitial porosity and bound water content data are available in the OTELo Research Data Repository (https://doi.org/10.24396/ORDAR-25). We thank CNRS-INSU and IODP-France for the funding of the shore-based measurements. We also thank David Billet, Philippe Rousselle, Maximilien Beuret and Géraldine Kitzinger from the LIEC laboratory in Nancy and Metz (France) where all CEC, exchangeable cation composition and soluble chloride content measurements were done. We greatly acknowledge Pierre Henry for fruitful discussions and Matt Ikari for the constructive review that greatly helped to improve the manuscript. We also thank Paradigm for support by providing Geolog software to visualize the log data.

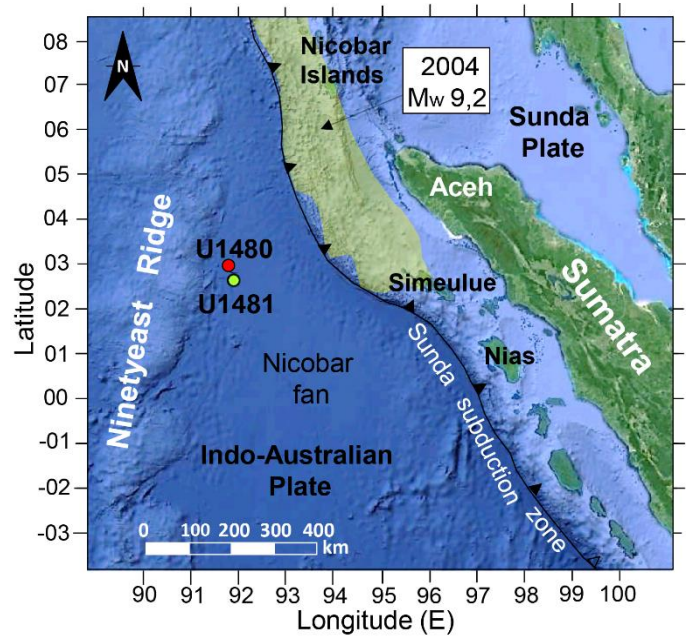

Figure 1. Map of Sites U1480 and U1481 and 2004 9.2 Mw Aceh-Andaman earthquake rupture zone (modified from Meltzner et al., 2012 and Hüpers et al., 2017). See McNeill et al., 2017a for seismic profiles corresponding to sampling locations.

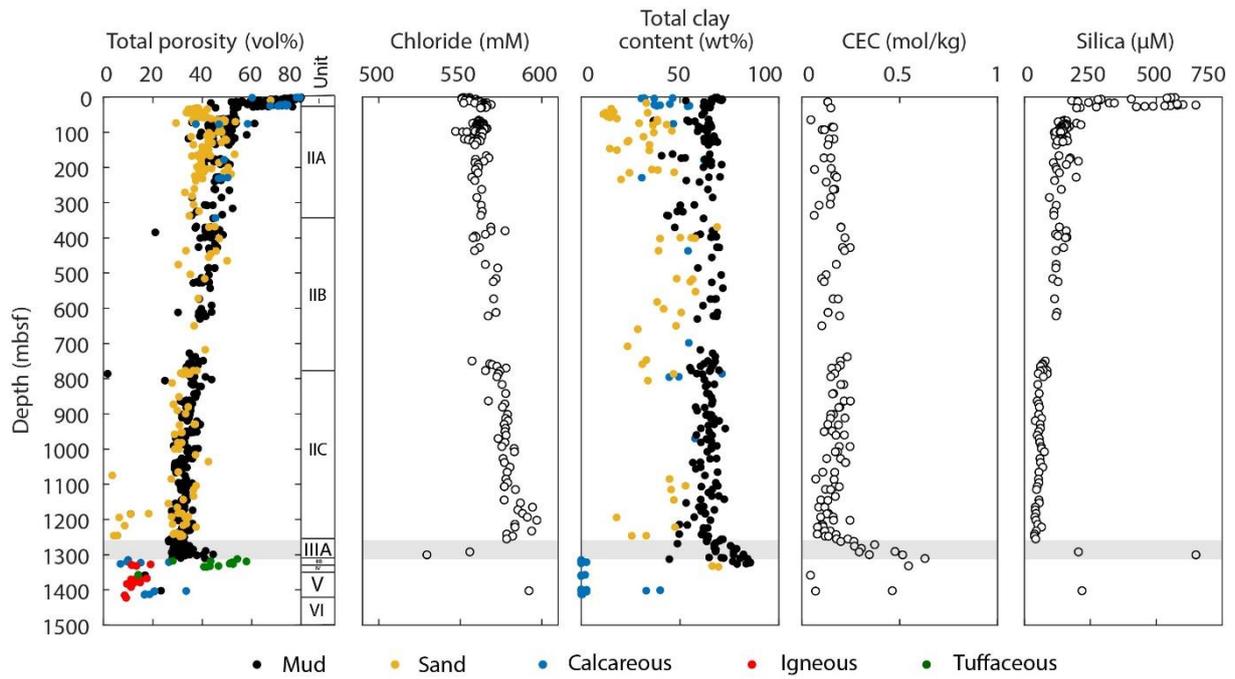

Figure 2. Physical, chemical and mineralogical data at Site U1480. Onboard MAD total porosity measured on mud, sand, calcareous, igneous and tuffaceous core samples shows a porosity anomaly (grey shaded zone) in mud samples from Unit III. This anomaly is correlated with a sharp decrease in pore water chloride content, a rise in total clay content, CEC and pore water silica content.

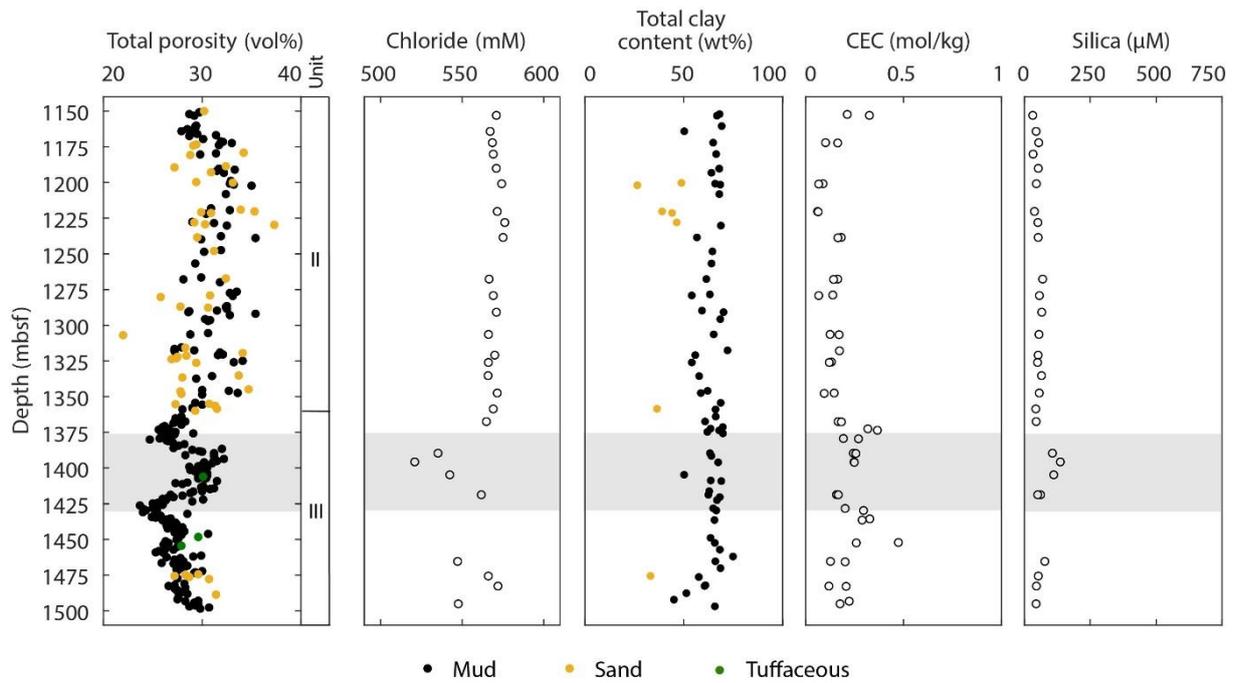

Figure 3. Physical, chemical and mineralogical data at Site U1481. Onboard MAD total porosity measured on mud, sand and tuffaceous core samples shows a porosity anomaly (grey shaded zone) in mud samples from Unit III. This anomaly is correlated with a sharp decrease in pore water chloride content and a rise in CEC and pore water silica content, which remains lower than at Site U1480.

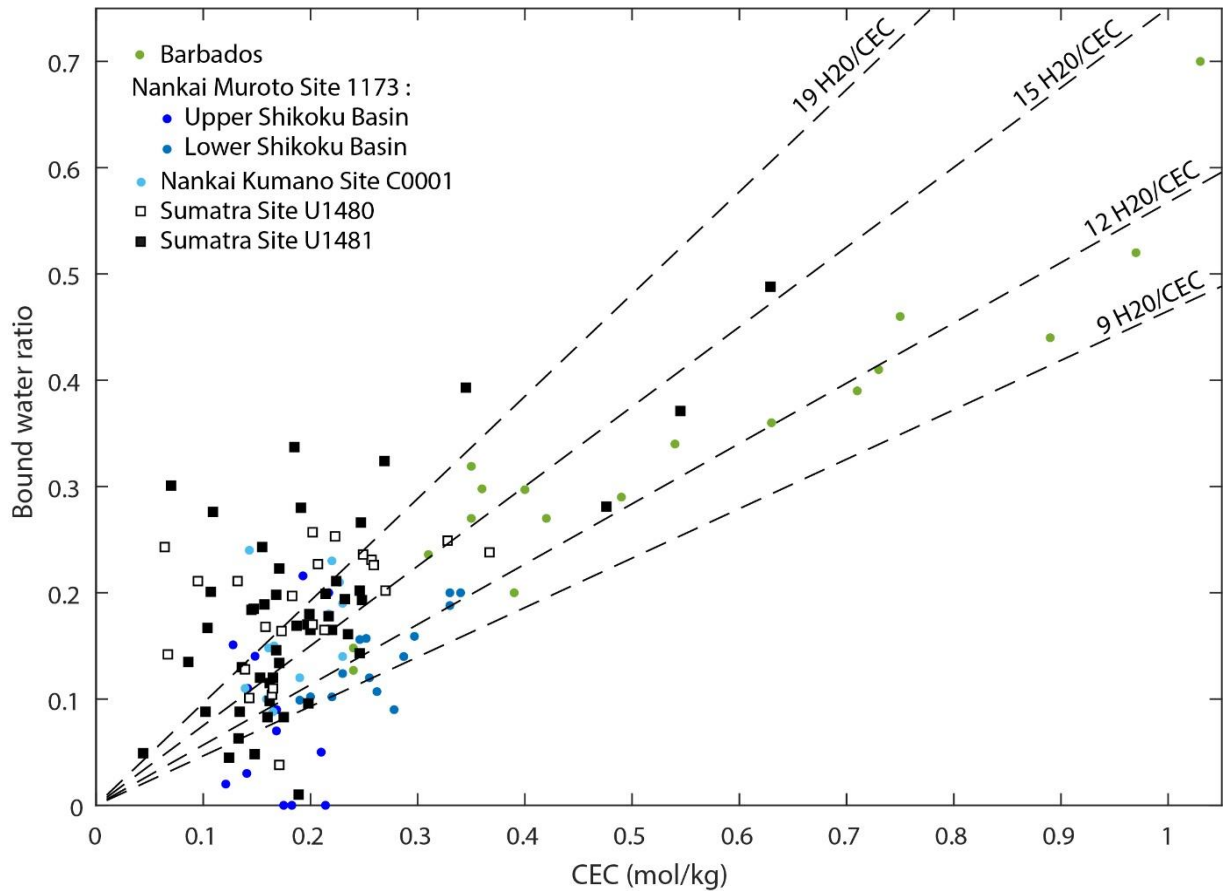

Figure 4. Bound water ratio versus Cation Exchange Capacity (CEC in mol/kg). The dashed-lines are from Henry and Bourlange (2004) and Conin et al. (2011) and correspond to an ideal smectite containing 9, 12, 15, 19 water molecules per cation charge. Data from Sumatra Site U1480 with relatively low clay content (Unit I and Subunit IIA) are not showed. Barbados data are from Henry, 1997, data from Nankai Muroto Lower Shikoku Basin (hemipelagites) and Upper Shikoku Basin (hemipelagites with more or less altered volcanic ash and biogenic silica) are from Henry & Bourlange, 2004 and data from Nankai Kumano Site C0001 (silty clay and hemipelagites from slope apron and accreted fine-grained terrigenous material and hemipelagites) are from Conin et al., 2011.

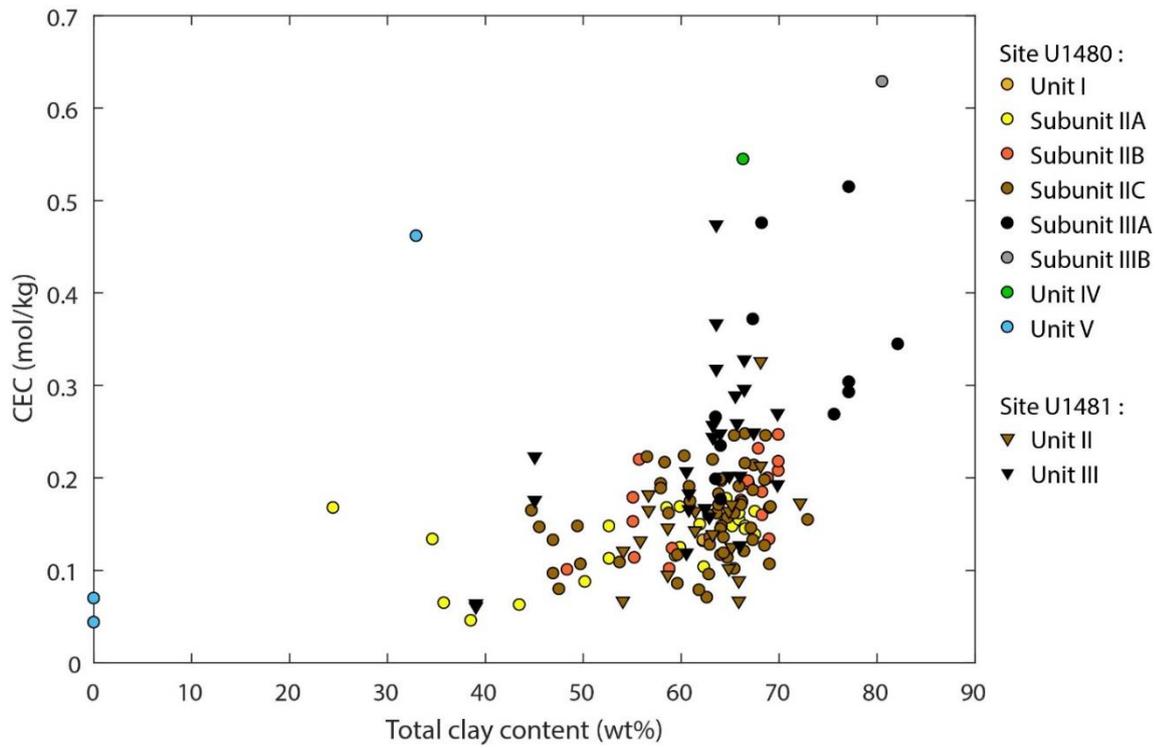

Figure 5. Cation Exchange Capacity (CEC in mol/kg) versus total clay content (normalized abundance weight percent) for samples of different lithology from Unit I to Unit V at Site U1480 and Unit II to Unit III at Site U1481.

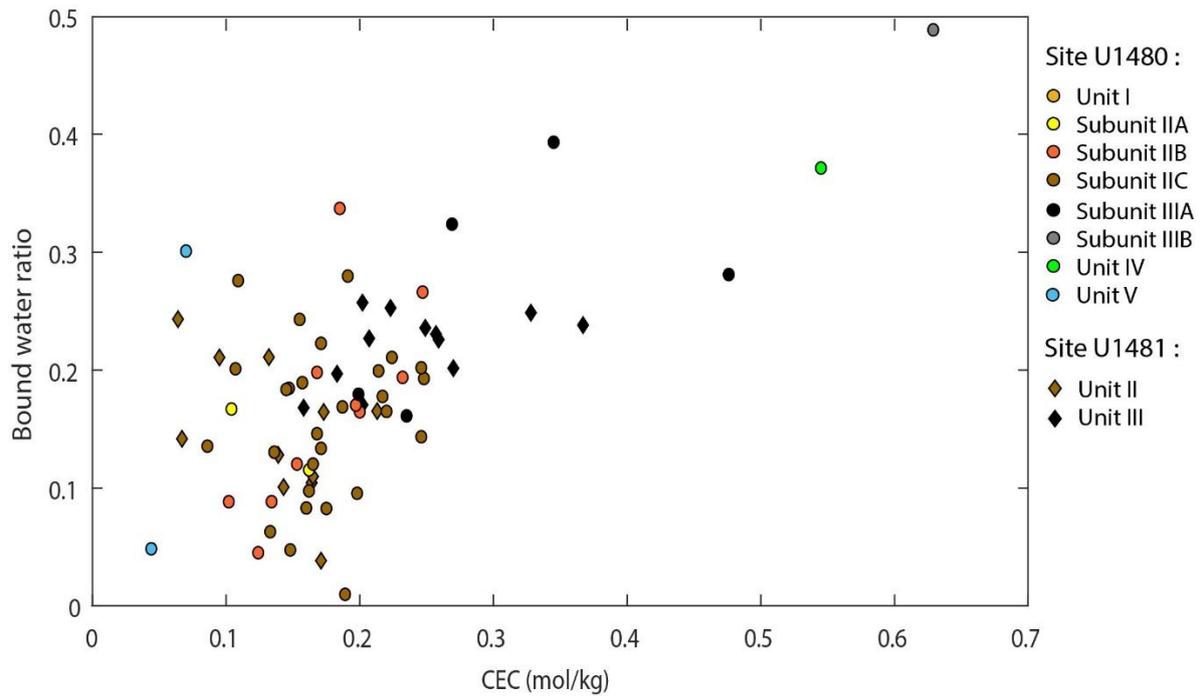

Figure 6. Bound water ratio versus Cation Exchange Capacity (CEC in mol/kg) for Sites U1480 and U1481 per unit.

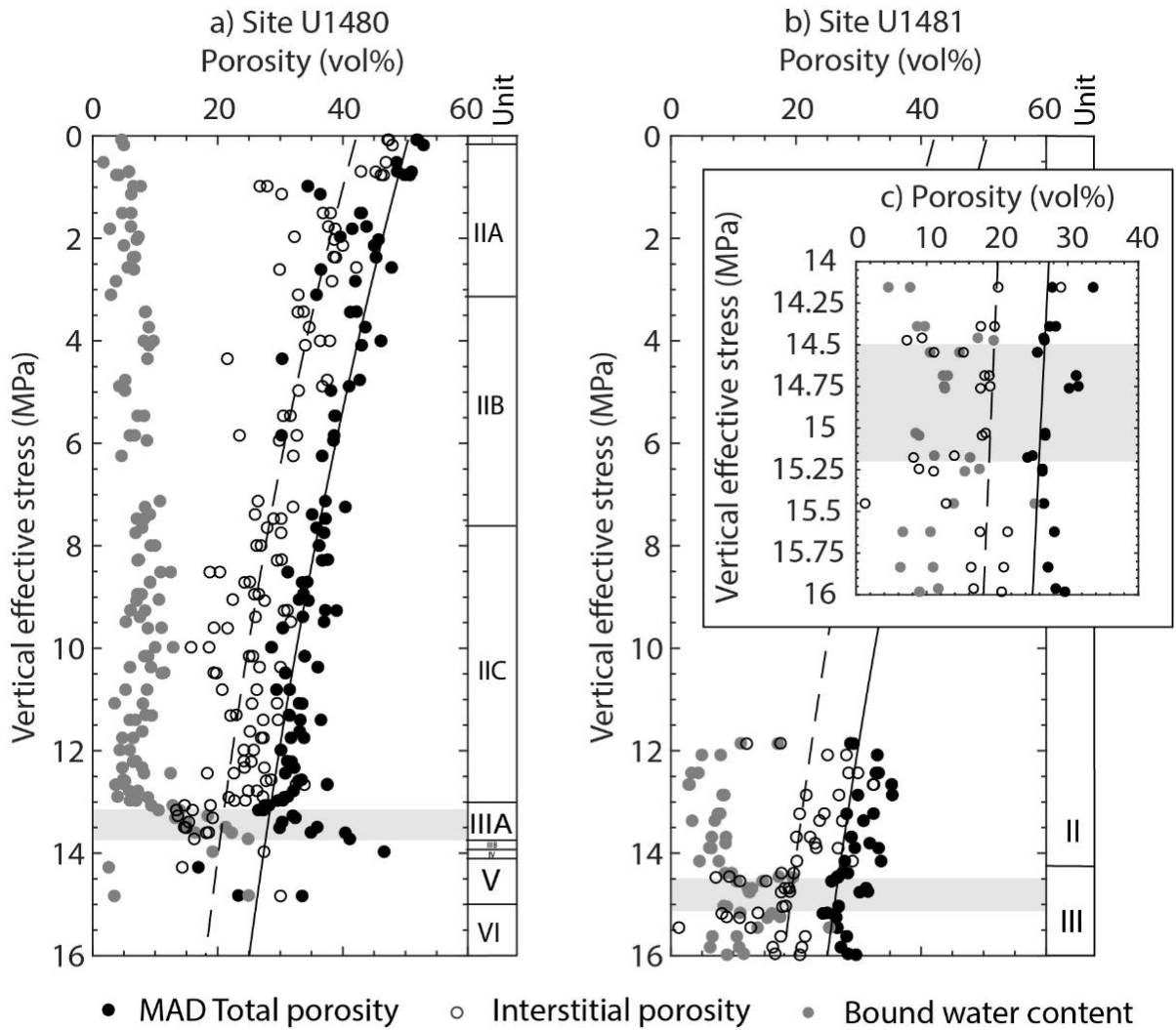

Figure 7. Total porosity, interstitial porosity and bound water content versus vertical effective stress at a) Site U1480 and b) Site U1481, with c) zoom in of the anomalously high total porosity zone at Site U1481. Bound water content is estimated from MAD total porosity and CEC data using $n = 15$ H$_2$0/CEC. The continuous line is the total porosity – effective stress curve $\phi_t = 50.6e^{-0.044\,\sigma'_v}$ ($R^2 = 0.7410$) calculated for MAD mud samples of Unit II at Site U1480. The discontinuous line is the interstitial porosity – effective stress curve $\phi_i = 42.2e^{-0.053\,\sigma'_v}$ ($R^2 = 0.6288$) calculated for the 118 samples analyzed for CEC at Site U1480. The grey shaded zone at each site represents the clay-rich zone of anomalously high total porosity.

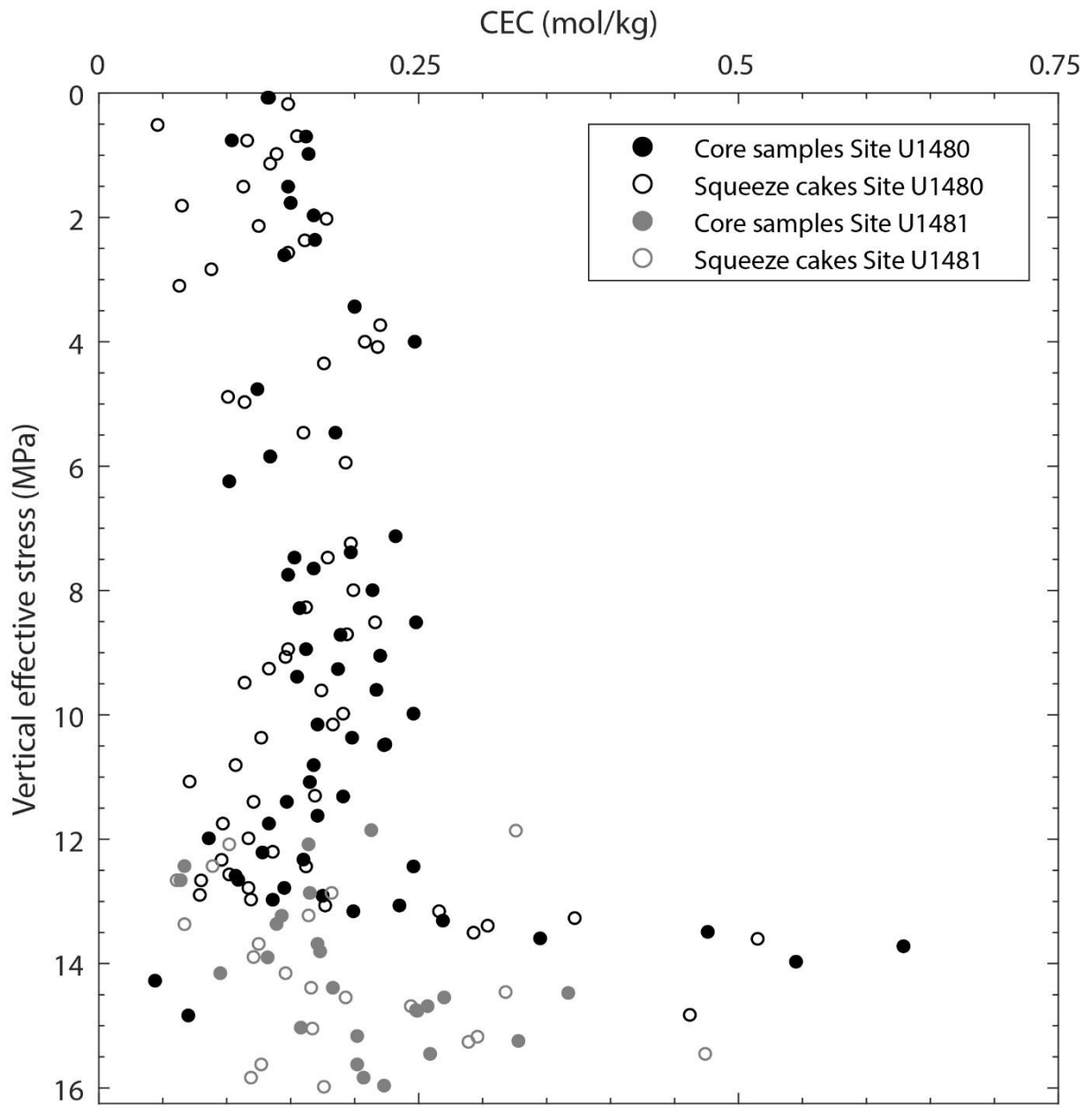

Figure 8. Cation Exchange Capacity measured on core samples and squeeze cakes at Sites U1480 and U1481 versus vertical effective stress.

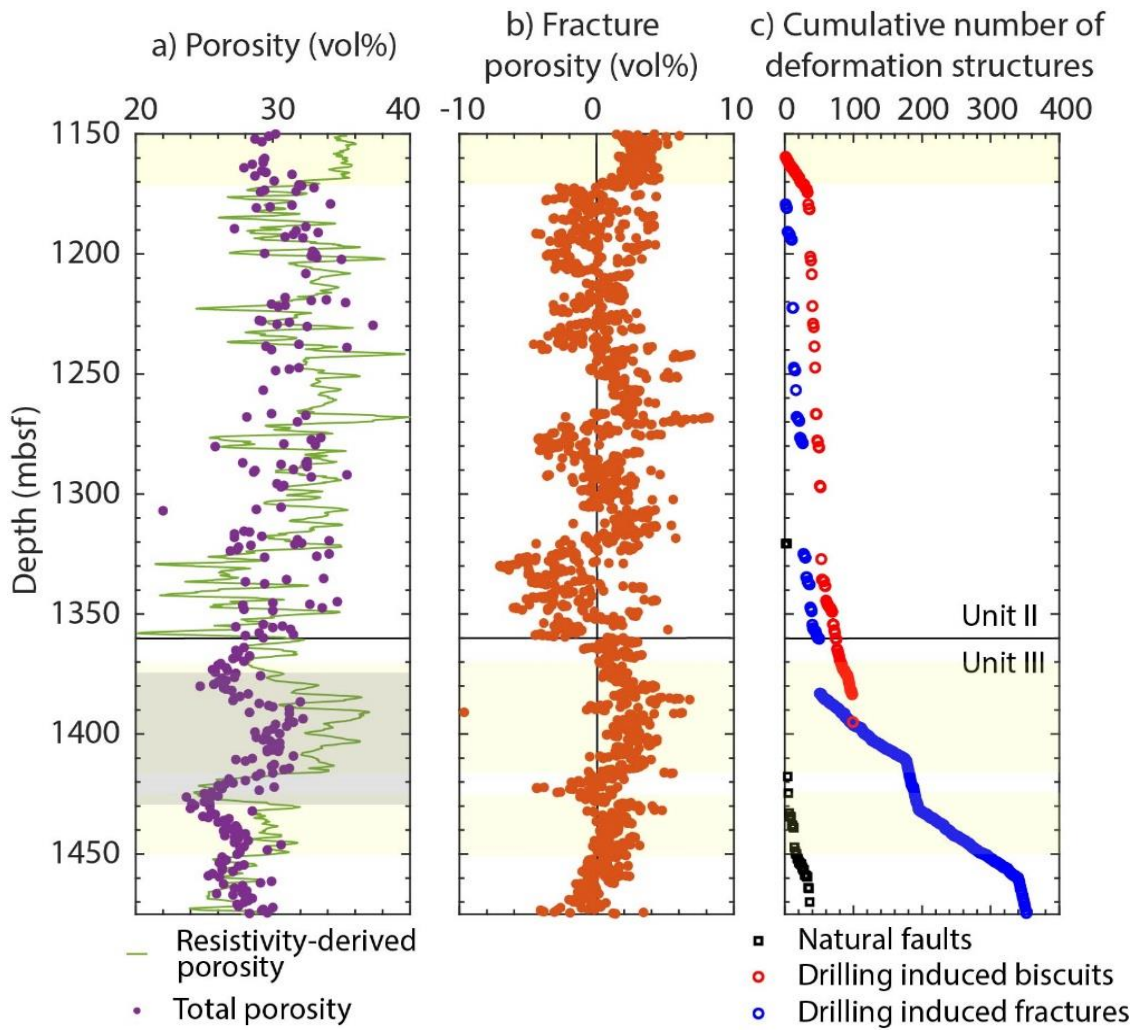

Figure 9. a) Comparison of MAD total porosity measured on cores at Site U1481 and porosity calculated from resistivity log at Site U1481 (sliding average for 10 resistivity values). b) Fracture porosity calculated from formation bulk conductivity, interstitial fluid conductivity and the conductivity of rock fragments at Site U1481. c) Cumulative number of faults, fractures and drilling biscuits observed on cores at Site U1481 (McNeill et al., 2017a). Yellow shaded zones ranging from ~1150 to 1170 mbsf, ~1370 to 1415 mbsf and ~1425 to 1450 mbsf correspond to areas where the discrepancy between resistivity-porosity and MAD total porosity is substantial. The grey shaded zone represents the clay-rich zone of anomalously high total porosity.

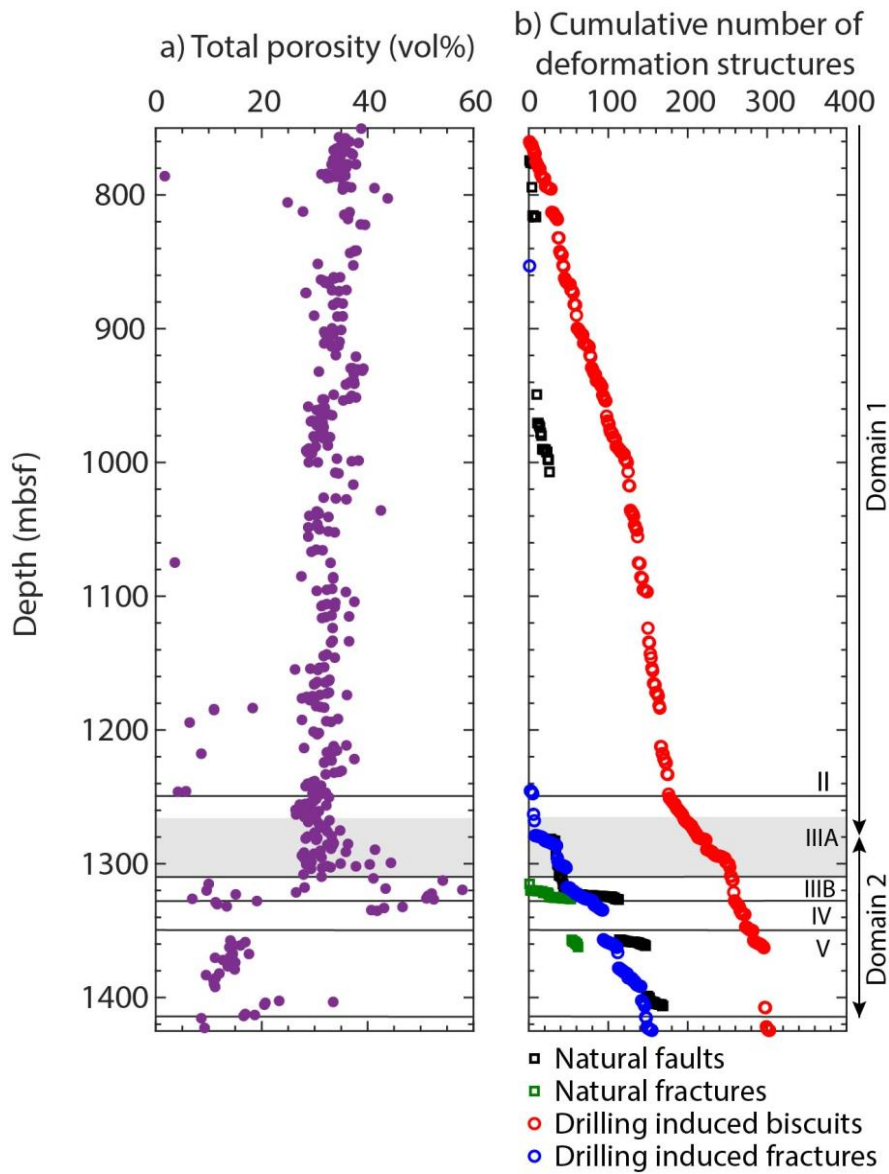

Figure 10. Zoom in on Site U1480 a) MAD total porosity measured on core samples and b) cumulative number of natural and drilling induced deformation structures observed on cores (McNeill et al., 2017a). The grey shaded zone represents the clay-rich zone of anomalously high total porosity.